\documentclass{article}

\usepackage{arxiv}

\usepackage[utf8]{inputenc} 
\usepackage[T1]{fontenc}    
\usepackage{hyperref}       
\usepackage{url}            
\usepackage{booktabs}       
\usepackage{amsfonts}       
\usepackage{nicefrac}       
\usepackage{microtype}      
\usepackage{lipsum}
\usepackage{amssymb}
\usepackage{amsfonts}
\usepackage{graphicx}
\graphicspath{ {figure/} }
\usepackage{subcaption}
\usepackage{epsfig}
\usepackage{float}
\usepackage{amsmath}
\usepackage[utf8]{inputenc}
\usepackage{amssymb}
\usepackage{nomencl}
\makenomenclature
\usepackage{hyperref}
\hypersetup{
    colorlinks,
    citecolor=black,
    filecolor=black,
    linkcolor=black,
    urlcolor=black
}
\usepackage{graphicx}
\graphicspath{ {./images/} }

\title{Eulerian formulation of the constitutive relation for an electro-magneto-elastic material class}

\author{
Deepak Kumar \\
  Department of Mechanical Engineering\\
  Maulana Azad National Institute of Technology Bhopal\\
  Madhya Pradesh, India 462003 \\
  \texttt{deepakkumaromre@manit.ac.in} \\
}

\begin{document}
\maketitle
\begin{abstract}
A novel class of electro-magneto-elastic (EME) materials comprise electro-active and magneto-active particles in the polymer matrix that change their elastic behavior with an applied electromagnetic field. The material response for such a material class is usually formulated in terms of Lagrangian strain tensor along with Lagrangian electromagnetic field vectors as “pushed forward” to the current configuration. This letter article presents a novel formulation of an electro-magneto-elasticity in terms of an Eulerian strain tensor and Eulerian electromagnetic field vectors referring to the current configuration. Such an Eulerian formulation is often favorable from both theoretical and computational standpoints, which avoids the “pushed forward” operation to get the current configuration. Additionally, an exercise to deduce the constitutive relation for an EME  material class available in the existing literature from the newly proposed relation is also illustrated.
\end{abstract}


From experience, we know that two bodies of the same `shape' and dimensions, with the same distribution of external load but made of different materials, will behave differently  \cite{kumar2021modeling,khurana2021effect,sharma2021finite}. Connecting the `deformation' with the stresses became necessary in such conditions \cite{kumar2022constitutive,kumar2021novel}. Generally, such connections are the simplified mathematical models relating the `deformation' with the `external loads', known as the constitutive relations for any material class \cite{gour2022constitutive, gour2022constitutive1, behera2021modeling}. Such relations primarily aim to represent approximately the material behavior without intending to model this behavior in all the possible external conditions. In this regard, to track the changes in elastic behavior of a smart elastomer responding to an applied electromagnetic field, another set of constitutive relations must be derived from the theory of electro-magneto-elasticity \cite{kumar2019electro, kumar2020universal}. In such relations, the material response is usually formulated using two commonly known Lagrangean and Eulerian strain tensors. Lagrangean and Eulerian strain tensors are generally related through the “pushed forward” and “pushed backward” operations. The “pushed forward” operation is a process of taking a Lagrangian quantity and writing its Eularian analog and vice-versa for reverse. The necessity of using Eulerian-based relations is often favorable from both theoretical and computational standpoints, which avoid the “pushed forward” operation to get the current configuration. In the reference configuration of an EME material class, the electro-active and magneto-active particles are not aligned. Rather, they are random. So, upon applying an electric field, magnetic field, or both together, the individual particles are now aligned in the current configuration, and the defined stress tensor will have a physical sense. But, when we convert all the variables to the reference configuration, that configuration becomes a pseudo configuration, not the actual one. So, to avoid such confusion, we propose an Eulerian approach to get the direct physical sense and the correct measure of the stresses, including all the kinematically dependent variables.

In the current scenario, various researchers, especially the material experimentalist and material modelers community, are very much interested in characterizing and segregating the various newly developed smart material classes, which fall under the category of non-elastic material polymers \cite{kumar2018instability, kumar2022modeling}. Such non-elastic material polymers are smart polymers in which different coupled phenomena like electromagnetostriction \cite{kumar2019electro} and electro-magneto-viscoelastic deformation \cite{kumar2021novel,behera2022constitutive} can be observed with an electromagnetic field of application. In view of the works on the electro-magneto-elasticity, a limited number of works, especially classical works \cite{alblas1974electro,bracke1981broadband,kovetz2000electromagnetic}, exist in the literature. Later, few authors \cite{carman1995micro,aboudi2001micromechanical} had explored the advancement in electro-magneto-elasticity by developing various micromechanical methods investigating the deformation response of the smart elastomers. In the same connection, the recent works \cite{kumar2019electro,kumar2020universal,kumar2022modeling,behera2022constitutive,khurana2021nonlinear,khurana2022static} also extended the novel advancements in electro-magneto-elasticity in obtaining the constitutive universal relations \cite{kumar2019electro,kumar2020universal} and analysing the smart actuator response with filler particles \cite{kumar2022modeling,khurana2021nonlinear,khurana2022static} including the damage-induced stress softening effect \cite{behera2022constitutive}. 

The existing works on electro-magneto-elasticity discussed above motivated us to develop novel constitutive equations investigating the deformation response of a novel electro-magneto-elastic (EME) material class. Inspired by them, the current study aims to present a direct Eulerian formulation of the constitutive relation for a novel EME material class. To the best of our knowledge, such an explicit Eulerian formulation of the constitutive relation in electro-magneto-elasticity is a first of its kind. \\

In the next, a significant contribution of this paper is formulated by presenting a theoretical modeling framework to develop a direct Eulerian formulation of the constitutive relation for a novel electro-magneto-elastic (EME) material class. Consider an EME solid continua occupying an initial stress-free configuration $\beta_0$ at time $t_0$ in the absence of any electric $\textbf{E}$, magnetic $\textbf{H}$, or mechanical loading. Let $\textbf{F}$ denote the deformation gradient of the initial stress-free configuration $\beta_0$ transformation to the current configuration $\beta$ at time $t$ subjected to any electric, magnetic, or mechanical loading condition. At the same time, the electro-magnetic field variables are governed by the following equations referred to, for example, \cite{kovetz2000electromagnetic} as 
\begin{align}\label{eqn:1}
\begin{split}
\mathrm{curl} \  \textbf{E}=0, \quad \mathrm{div} \  \textbf{D}=0, \quad \mathrm{curl} \  \textbf{H}=0, \quad \mathrm{div} \  \textbf{B}=0,
\end{split}
\end{align}
where new variables $\textbf{D}$ and $\textbf{B}$ denote the electric induction (or electric displacement) and magnetic induction (or magnetic displacement) vectors, respectively. In addition, curl and div are the curl and divergence operators with respect to the current spacial vector. For a given continua, these new electromagnetic field variables are defined as \cite{chelkowski1980dielectric,callen1968magnetostriction}
\begin{align}\label{eqn:2}
\begin{split}
\textbf{D}=\epsilon_0 \textbf{E}+\textbf{P}, \quad \textbf{B}=\mu_0 (\textbf{H}+\textbf{M}),
\end{split}
\end{align}
where $\textbf{P}$ and $\textbf{M}$ represent the polarization and magnetization densities of the material having $\epsilon_0$ and $\mu_0$ electric permittivity and magnetic permeability
of free space, respectively. If $\textbf{E}$ and $\textbf{H}$ are taken as the independent field variable then the above equation (\ref{eqn:2}) gives the other electric variable $\textbf{D}$ (respectively $\textbf{P}$) when $\textbf{P}$
(respectively $\textbf{D}$) is prescribed by a constitutive equation, along with the magnetic variable $\textbf{B}$ (respectively $\textbf{M}$) when $\textbf{M}$
(respectively $\textbf{B}$) is prescribed. For a given EME continua subjected to an electromagnetic field deformation case, the Lagrangian (or “pull back” from $\beta$ to $\beta_0$) version of characteristic applied electric and magnetic field directions in the reference configuration are given by
\begin{align}\label{eqn:3}
\begin{split}
\textbf{E}^l= \textbf{F}^T \textbf{E}, \quad \textbf{D}^l= \textbf{F}^{-1} \textbf{D}, \quad \textbf{P}^l=  \textbf{F}^{-1} \textbf{P}, \quad 
\textbf{H}^l= \textbf{F}^T \textbf{H}, \quad \textbf{B}^l= \textbf{F}^{-1} \textbf{B}, \quad \textbf{M}^l=  \textbf{F}^{-1} \textbf{M}. 
\end{split}
\end{align}
The Eulerian (or “pull forward” from $\beta_0$ to $\beta$) equivalent  version of electric and magnetic field directions in the current configuration are now obtained by differentiation of equations  (\ref{eqn:3})  with respect to time as 
\begin{align}\label{eqn:4}
\begin{split}
 \dot {\textbf{E}}^l= \textbf{L}^T \textbf{E}^l, \quad \dot {\textbf{D}}^l=  \textbf{L}^T\textbf{D}^l, \quad \dot {\textbf{P}}^l= \textbf{L}^T\textbf{P}^l, \quad 
 \dot {\textbf{H}}^l= \textbf{L}^T \textbf{H}^l, \quad \dot {\textbf{B}}^l= \textbf{L}^T\textbf{B}^l, \quad \dot {\textbf{M}}^l= \textbf{L}^T\textbf{M}^l,
\end{split}
\end{align}
where $\textbf{L}=\mathrm{grad} \textbf{v}= \dot {\textbf{F}} \textbf{F}^{-1}$ represents the velocity gradient tensor. Note that the electric $\textbf{E}^l$ and magnetic $\textbf{H}^l$ may be replaced by $-\textbf{E}^l$ and  $-\textbf{H}^l$ for reverse direction, which is equivalently defined the characteristic anisotropy. To get rid of such non-uniqueness, the electric $\textbf{e}$ and magnetic  $\textbf{h}$ tensors may be defined as
\begin{align}\label{eqn:5}
\begin{split}
\textbf{e}=\textbf{E}^l \otimes \textbf{E}^l, \quad \textbf{h}=\textbf{H}^l \otimes \textbf{H}^l.
\end{split}
\end{align}
Using equations (\ref{eqn:3}) and (\ref{eqn:4}), evolution equations for the above electric $\textbf{e}$ and magnetic  $\textbf{h}$ tensors are given by
\begin{align}\label{eqn:6}
\begin{split}
\dot{\textbf{e}}= \textbf{L} \textbf{e}+ \textbf{e} \textbf{L}^T, \quad  \dot{\textbf{h}}= \textbf{L} \textbf{h}+ \textbf{h} \textbf{L}^T.
\end{split}
\end{align}
At the same time,  mechanical deformation in the Eulerian
description that is an evolution equation similar to (\ref{eqn:6}) for a defined left Cauchy–Green deformation $\textbf{b}=\textbf{F}\textbf{F}^T$ in the Lagrangian description is obtained as
\begin{align}\label{eqn:7}
\begin{split}
\dot{\textbf{b}}= \dot{\textbf{F}} \textbf{F}^T+ \textbf{F} \dot{\textbf{F}}^T
=\dot{\textbf{F}} \left( \textbf{F}^{-1} \textbf{F}\right) \textbf{F}^T+\textbf{F} \left( \textbf{F}^{-1} \textbf{F}\right)^T \dot{\textbf{F}}^T \quad 
=(\dot{\textbf{F}} \textbf{F}^{-1}) (\textbf{F} \textbf{F}^T) 
+(\textbf{F} \textbf{F}^T) (\textbf{F}^{-T}\dot{\textbf{F}}^T) 
=\textbf{L}\textbf{b}+\textbf{b}\textbf{L}^T.
\end{split}
\end{align}
Hence, a kinematic tensor $\dot{\textbf{b}}$ for the mechanical deformation and the kinematic electric tensor $\dot{\textbf{e}}$  along with a magnetic part $\dot{\textbf{h}}$ for an electromagnetic deformation are described in the Eulerian description via evolution equations (\ref{eqn:6}) and (\ref{eqn:7}).  Such a set of primarily defined kinematic field variables $(\dot{\textbf{b}}, \dot{\textbf{e}}, \dot{\textbf{h}})$ in the Eulerian description are now get ready to deduce the constitutive relations for an EME material class. \\

To develop the constitutive relations for an EME material class, a purely EME deformation is assumed here by ignoring an internal dissipation of the material. The corresponding thermodynamically consistent Clausius–Duhem inequality-based internal dissipation $D_{int}$ vanishing for the defined EME system of continua with the defined effective conduction current $\textbf{J}_e$, effective electric current $\textbf{E}_e=\textbf{E}+\textbf{v} \times \textbf{B}$, and effective magnetization $\textbf{M}_e= \mu_0^{-1}\textbf{M}+\textbf{v} \times \textbf{P}$  (see \cite{bustamante2007mathematical,kumar2019mathematical,kumar2020universala}, for details) is given by 
\begin{align}\label{eqn:8}
\begin{split}
D_{int}=\textbf{T} : \textbf{d}- \varrho \dot{w} +\textbf{J}_e.\textbf{E}_e- \dot{\textbf{E}}_e \textbf{P}-\textbf{M}_e. \dot{\textbf{B}} =0,
\end{split}
\end{align}
where $\textbf{T}$ represents the Cauchy stress tensor, $\textbf{d}=(\textbf{L}+\textbf{L}^T)/2$ denotes the symmetric part of the deformation rate tensor, and $\varrho$ and $w$ are the current mass density along with the specific free energy per unit mass. By targeting the Eulerian description of the constitutive relations for an EME material class, the specific free energy $w(\dot{\textbf{b}}, \dot{\textbf{e}}, \dot{\textbf{h}})$ is assumed as a function of the defined kinematic variables. Consequently,
the rate of the free energy is given by
\begin{align}\label{eqn:9}
\begin{split} 
\dot{w}(\dot{\textbf{b}}, \dot{\textbf{e}}, \dot{\textbf{h}})= \dfrac{\partial w}{\partial \textbf{b}}: \dot{\textbf{b}}+\dfrac{\partial w}{\partial \textbf{e}}: \dot{\textbf{e}}+\dfrac{\partial w}{\partial \textbf{h}}: \dot{\textbf{h}}.
\end{split}
\end{align}
Substituting the equations (\ref{eqn:6}) and (\ref{eqn:7}) in the above equation (\ref{eqn:9}), the rate of the free energy expression is re-written as
\begin{align}\label{eqn:10}
\begin{split} 
\dot{w}(\dot{\textbf{b}}, \dot{\textbf{e}}, \dot{\textbf{h}})= 2 (\textbf{Z}:\textbf{L}),
\end{split}
\end{align}
where 
\begin{align}\label{eqn:11}
\begin{split} 
\textbf{Z}=\dfrac{\partial w}{\partial \textbf{b}} \textbf{b}+\dfrac{\partial w}{\partial \textbf{e}} \textbf{e}+\dfrac{\partial w}{\partial \textbf{h}} \textbf{h}.
\end{split}
\end{align}
With the use of equation (\ref{eqn:11}) on (\ref{eqn:8}), the internal dissipation expression vanishing for the defined EME
system of continua is re-written as
\begin{align}\label{eqn:12}
\begin{split}
D_{int}=\textbf{T} : \textbf{d}- \varrho \left( \textbf{Z}+\textbf{Z}^T\right): \textbf{d}- \varrho \left( \textbf{Z}+\textbf{Z}^T\right): \textbf{w} \quad 
+\textbf{J}_e.\textbf{E}_e- \dot{\textbf{E}}_e \textbf{P}-\textbf{M}_e. \dot{\textbf{B}} =0,
\end{split}
\end{align}
where $\textbf{w}=(\textbf{L}-\textbf{L}^T)/2$ denotes the anti-symmetric part of the deformation rate tensor, known as a spin tensor.
Assuming the components of $\textbf{d}$ and $\textbf{w}$ as independent, a given tensor $\textbf{Z}$ is to be needed as $\textbf{Z}=\textbf{Z}^T$. Using the equations (\ref{eqn:11}) in (\ref{eqn:12}) that is expected to hold for real materials at all times and at every fixed point in space for a certain class of admissible thermodynamic processes, the constitutive equations for an EME material class are obtained as
\begin{align}\label{eqn:13}
\begin{split}
\textbf{T}= 2 \varrho \textbf{Z}, \quad \textbf{P}=-\varrho \dfrac{\partial w}{\partial \textbf{E}_e}, \quad \textbf{M}_e=-\varrho \dfrac{\partial w}{\partial \textbf{B}}.
\end{split}
\end{align}
The above equation represents an alternative form of the constitutive relations formulated directly with an Eulerian description for an EME material class in line with the literature \cite{kumar2019electro, kumar2020universal, kumar2018instability, kumar2022modeling}. Utilizing the above newly proposed relation (\ref{eqn:13}), the upcoming section aimed to deduce the original form of the relations available in the existing literature. \\

At last, the original form of the constitutive relations existing in the literature is deduced from the newly developed constitutive relation (\ref{eqn:13}) in the previous section. By way of example, considering an electro-magneto-elasticity case with
an applied electric $\textbf{E}$ and magnetic $\textbf{H}$ fields, wherein the free energy $w(I_1, I_2, ... I_9)$ depending on nine invariants given by \cite{kumar2019electro, kumar2020universal}
\begin{align}\label{eqn:14}
\begin{split}
{I_1= \mathrm{tr} \textbf{b}},\quad {I_2= \dfrac{1}{2}[ (\mathrm{tr} \textbf{b})^2-\mathrm{tr} (\textbf{b}^2)]},\quad {I_3= \mathrm{det} \textbf{b}}\quad 
{I_4= \textbf{e}:\textbf{I}},\quad {I_5= \textbf{e}:\textbf{b}^{-1}}, \quad 
{I_6= \textbf{e}:\textbf{b}^{-2}}, \\
 {I_7=  \textbf{h}:\textbf{I}}, \quad
{I_8= \textbf{h}:\textbf{b}^{-1}}, \quad
{I_9= \textbf{h}:\textbf{b}^{-2}}.
\end{split}
\end{align}
The first three invariants are the principal ones that correspond to the mechanical deformation, while the last six
invariants correspond to the electromagnetic field effect. It is to be noted that in the Eulerian description, the invariants $I_4$ to $I_9$ linked with electromagnetic effects are generally functions of both deformation tensor $\textbf{b}$ and the corresponding field tensor either $\textbf{e}$ or $\textbf{h}$. The corresponding nontrivial derivatives of such invariants are obtained as
\begin{align}\label{eqn:15}
\begin{split}
\dfrac{\partial I_1}{\partial \textbf{b}}= \textbf{I}, \quad \dfrac{\partial I_2}{\partial \textbf{b}}= I_1 \textbf{I}-\textbf{b}, \quad \dfrac{\partial I_3}{\partial \textbf{b}}= I_3\textbf{b}^{-1}, \quad 
\dfrac{\partial I_4}{\partial \textbf{e}}= \textbf{I}, \quad \dfrac{\partial I_5}{\partial \textbf{e}}= \textbf{b}^{-1}, \quad \dfrac{\partial I_6}{\partial \textbf{e}}= \textbf{b}^{-2}, \\
\dfrac{\partial I_7}{\partial \textbf{h}}= \textbf{I}, \quad \dfrac{\partial I_8}{\partial \textbf{h}}= \textbf{b}^{-1}, \quad \dfrac{\partial I_9}{\partial \textbf{h}}= \textbf{b}^{-2}.
\end{split}
\end{align}
For a given invariants-dependent free energy function $w(I_1, I_2,...I_9)$ along with the equation (\ref{eqn:10}) yields the following expression 
\begin{align}\label{eqn:16}
\begin{split}
\dfrac{\partial w}{ \partial \textbf{b}}= \sum_{n=1}^{9} w_n \dfrac{\partial I_n}{ \partial \textbf{b}}, \quad   \dfrac{\partial w}{ \partial \textbf{e}}= \sum_{n=4}^{6}w_n \dfrac{\partial I_n}{ \partial \textbf{e}},\quad 
\dfrac{\partial w}{ \partial \textbf{h}}= \sum_{n=7}^{9}w_n \dfrac{\partial I_n}{ \partial \textbf{h}},
\end{split}
\end{align}
where $w_n=\dfrac{\partial w}{\partial I_n}$.
Substituting the corresponding nontrivial derivatives expressions from (\ref{eqn:15}) in (\ref{eqn:16}), the above expressions (\ref{eqn:16}) may be re-written as
\begin{align}\label{eqn:17}
\begin{split}
\dfrac{\partial w}{ \partial \textbf{b}}
=(w_1+w_2I_1)\textbf{I}-w_2\textbf{b}
+w_3I_3\textbf{b}^{-1}
- w_5 \textbf{b}^{-2} \textbf{e} 
- w_6 \textbf{b}^{-3} \textbf{e}- w_8 \textbf{b}^{-2} \textbf{h}
- w_9 \textbf{b}^{-3} \textbf{h}, \\
\dfrac{\partial w}{ \partial \textbf{e}}
=w_4\textbf{I}+w_5\textbf{b}^{-1}+w_6\textbf{b}^{-2},\quad 
\dfrac{\partial w}{ \partial \textbf{h}}
=w_7\textbf{I}+w_8\textbf{b}^{-1}+w_9\textbf{b}^{-2},
\end{split}
\end{align}
Further, substituting the equations (\ref{eqn:11}), (\ref{eqn:16}), and (\ref{eqn:17}) in a newly developed form of the constitutive relations (\ref{eqn:13}) formulated directly with an Eulerian description, we get
\begin{align}\label{eqn:18}
\begin{split}
\textbf{T}= 2 \varrho [ (w_1+w_2 I_1)\textbf{b}-w_2\textbf{b}^2+w_3I_3 \textbf{I}+w_4 \textbf{e} 
+w_5(\textbf{b}^{-1} \textbf{e}+\textbf{e}\textbf{b}^{-1} ) +w_6(\textbf{b}^{-2} \textbf{e}+\textbf{e}\textbf{b}^{-2} )\\
+w_8(\textbf{b}^{-1} \textbf{h}+\textbf{h}\textbf{b}^{-1} ) +w_9(\textbf{b}^{-2} \textbf{h}+\textbf{h}\textbf{b}^{-2} )
].
\end{split}
\end{align}
Finally, the above-obtained relation is an equivalent original form of the constitutive relation to evaluate the stress response compared with the existing literature \cite{kumar2019electro, kumar2020universal, kumar2018instability, kumar2022modeling}. \\

In the summary, a novel formulation of an electro-magneto-elasticity in terms of an Eulerian strain tensor and Eulerian electromagnetic field vectors referring to the current configuration is developed. To develop the same, a thermodynamically consistent continuum
mechanics-based modeling approach is adopted. A newly proposed relation (\ref{eqn:13}) is further exercised in obtaining the original form of the constitutive relation (\ref{eqn:18}) available in the existing literature. In short, it is emphasized in the current work that Eulerian-based relations are often favorable compared to Lagrangian from both theoretical and computational standpoints, which avoid the “pushed forward” operation to get the current configuration.

\section*{Acknowledgement}
The author is grateful to Dr. Somnath Sarangi, Associate Professor in the Department of Mechanical Engineering, IIT Patna, India, for the fruitful discussions on the topic.

\bibliographystyle{elsarticle-num}  
\bibliography{references}  


\end{document}